# Wideband Perfect Absorption Polarization Insensitive Reconfigurable Graphene Metasurface for THz Wireless Environment

Sasmita Dash
Computer Science Department
University of Cyprus
Nicosia, Cyprus
sasmitadash30@ieee.org

Christos Liaskos
Institute of Computer Science
Foundation of Research and Technology Hellas
Heraklion, Greece
cliaskos@ics.forth.gr

Ian F. Akyildiz
School of Electrical and Computer Engineering
Georgia Institute of Technology
Atlanta, USA
ian@ece.gatech.edu

Andreas Pitsillides
Computer Science Department
University of Cyprus
Nicosia, Cyprus
andreas.pitsillides@ucy.ac.cy

*Abstract*— **In this work, we investigated a simple structured graphene terahertz (THz) metasurface (MSF) with perfect absorption, wideband, polarization insensitive, oblique incidence insensitive and frequency reconfiguration. The graphene MSF structure is composed of a two-dimensional periodic array of graphene meta-atoms deposited on the silicon substrate terminated by a metal ground plane. The performance of the proposed MSF is numerically analyzed. An equivalent circuit model of the structure and its closed-form solution is introduced. The graphene MSF thin structure at 2.5 THz provides 100% of absorption with wide bandwidth, zero reflection and zero transmission at normal incidence in both transverse electric (TE) and transverse magnetic (TM) polarization. Under oblique incidence, the absorption is maintained at higher than 95%. Moreover, the graphene MSF structure has the advantage of frequency reconfiguration. The excellent absorption performance is maintained at all reconfigurable frequencies upon reconfiguration. The results reveal the effectiveness of the THz MSF with graphene meta-atoms, which can be promising for THz wireless environment.**

## I. Introduction

Owing to its two-dimensional (2D) nature and the unique properties at THz frequencies, graphene has attracted a lot of attention to THz research community. It has found its applications in designing miniaturized devices required for THz wireless communication [1-3]. The most important properties of graphene at THz frequency is the ability to support the propagation of the surface plasmon polariton (SPP) waves [4]. The graphene plasmons possess more confinement, low loss and good tunability. Furthermore, the conductivity of graphene is determined by the Fermi level and can be tuned by the electric field, chemical doping and gate voltage. Due to the properties of conductivity function and plasmonic effect at THz frequencies, graphene-based devices enables high miniaturization and reconfiguration [1], [5]. The unique feature makes graphene a promising material for tunable devices, especially in the THz range. The strong optical response arising from the graphene surface plasmons also enables novel metamaterial [6]. Because of its tunable properties, graphene can be a promising candidate for reconfigurable THz MSF.

In recent years, MSF has gained significant attention due to its ability to control the wave propagation by patterning planar subwavelength structure. Hypersurface (HSF), a programmable MSF, design exhibits a high degree of MSF pattern customization and allows for a high degree of tunability [7]. HSF empowered wireless environment enables the optimization of the propagation factor between wireless devices [7], [8]. Copper is the most widely used metal in radio frequency and microwave frequency range. But, the design of THz MSF faces many challenges, that range from the technological limitation of microfabrication to requirement for consideration of the electromagnetic interaction at nanoscale. To achieve the reconfigurability, switching mechanism approaches have been used in metal MSF structure. The incorporation of switches into the MSF structure increases the complexity of the THz MSF structure.

In this work, we propose a simple structure of the graphene MSF at low THz frequencies. This work numerically investigates the absorption performance of graphene MSF structure. The performance of the graphene MSF is analyzed under normal and oblique incidences in both TE and TM polarization. Moreover, the impact of variation of the Fermi level on the graphene MSF is studied, which provide reconfigurable MSF. An equivalent circuit modeling of the structure and its closed-form solution is also introduced. The MSF with graphene meta-atom designs could act as a smart environment for THz communications.

## II. Design and Circuit Modelling of Graphene MSF

A smart THz MSF using graphene meta-atom is presented in this work. The graphene MSF structure is composed of a 2D

periodic array of graphene elements deposited on silicon substrate terminated by a metal ground plane. Fig. 1 (a) and (b) shows the cross-sectional view and the 3D view of the proposed graphene MSF structure respectively. An equivalent circuit model of the structure is presented in Fig. 1(c).

In the literature, the periodic array of graphene meta-atoms on a dielectric layer over a metallic plane has been discussed [9-13]. In comparison to the previous works, a simple graphene MSF structure at THz band is used in this work, for a thin, excellent, wideband, wide-angle incidence, polarization-insensitive, and reconfigurable absorber.

The periodic array consists of the graphene square patches with dimension of $d \times d$, set along the x- and y-directions with period $P$. The operational frequency is considered as $f = 2.5$ THz. To achieve thin absorber, the thickness of the substrate is considered as $h = \lambda/13$. The thickness $t$ of 0.3 μm is chosen for the metallic plane. The patch width and the period are $d = \lambda/14$ and $P = \lambda/10$ respectively. All dimensions of the proposed structure are in μm units. We assumed the electron mobility of graphene $\mu_g = 2000$ cm$^2$/Vs, the electron relaxation time $\tau = 0.1$ ps, and temperature $T = 300$ K in this work. The properties and dimensions of the graphene MSF structure are chosen for the design of thin, perfect, wideband absorber at THz frequencies.

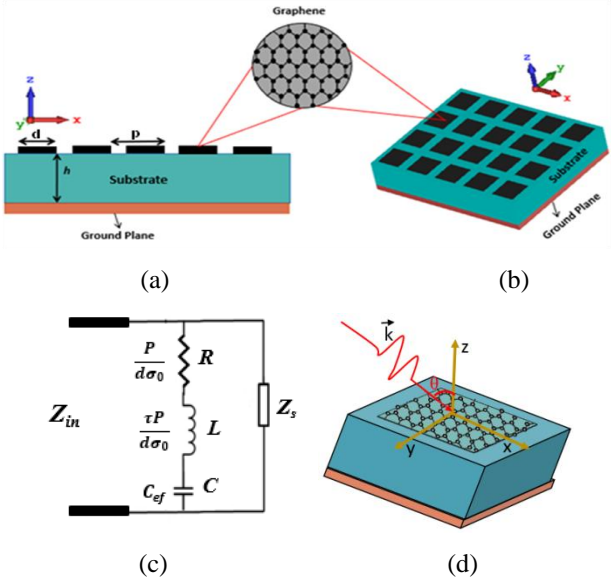

Fig. 1. Schematic of the graphene MSF (a) Cross-sectional view and (b) 3D view, (c) Equivalent circuit of graphene MSF structure, and (d) it's unit cell.

The graphene elements are modeled as infinitesimally thin sheets with the surface impedance $Z = 1/\sigma(\omega)$, where $\sigma(\omega)$ is the frequency-dependent complex conductivity of graphene. The conductivity of graphene is described by the Kubo formula [14], which is the contribution from both intraband transition and interband transition. The graphene conductivity is dependent on angular frequency $\omega$, electron relaxation time $\tau$, temperature $T$, and chemical potential $\mu_c$.

$$\sigma_g(\omega, \mu_c, \tau, T) = \sigma_{intra} + \sigma_{inter}$$

$$\sigma_g(\omega,\mu_c,\tau,T) = -j\frac{e^2 \kappa_B T}{\pi \hbar^2 (\omega - j\tau^{-1})} \left[ \frac{\mu_c}{\kappa_B T} + 2\ln\left(e^{-\mu_c/\kappa_B T} + 1\right) \right] - \frac{je^2}{4\pi\hbar} \ln\left[ \frac{2|\mu_c| - \hbar(\omega - j\tau^{-1})}{2|\mu_c| + \hbar(\omega - j\tau^{-1})} \right] \quad (1)$$

where $e$ is the electron charge, $k_B$ is the Boltzmann's constant, $\hbar$ is the reduced Planck's constant, $T$ is the temperature, $\mu_c$ is the chemical potential (Fermi energy), $\tau$ is the relaxation time. 1$^{st}$ and 2$^{nd}$ term of the above equation represents intraband and interband contribution for graphene conductivity. In the low THz band, the intraband conductivity dominates over interband conductivity. For $\mu_c \gg k_B T$, the conductivity of graphene $\sigma_g$ at the low THz band can be expressed as

$$\sigma_g = \frac{e^2 \mu_c \tau}{\pi \hbar^2} \frac{1}{(1 + j\omega\tau)} \quad (2)$$

Above equation in the Drude form can be calculated as,

$$\sigma_g = \frac{\sigma_0}{(1 + j\omega\tau)} \quad (3)$$

where $\sigma_0 = \frac{e^2 \mu_c \tau}{\pi \hbar^2}$

The surface impedance of the graphene patch array can be approximated as [15]

$$Z_g = \frac{P}{(P-s)\sigma_g} - j\frac{\pi}{\omega\varepsilon_0(\varepsilon_r + 1)P \ln\{\csc(\pi s/2P)\}} \quad (4)$$

where $P$ is the period of graphene array, $s$ is the space between the graphene elements, $\varepsilon_r$ is the relative permittivity of the substrate. Eq. (4) can be rewritten as

$$Z_g = \frac{P}{d\sigma_g} - j\frac{1}{\omega C_{ef}}$$

(5)

where $P-s = d$ is the width of graphene elements and $C_{ef} = (1/\pi) \varepsilon_0(\varepsilon_r+1) P \ln \{\csc (\pi s/2P)\}$ is the effective capacitance arises from spaces between graphene elements.

Using Eq. (3), Eq. (5) becomes

$$Z_g = \frac{P}{d\sigma_0} + j\left[\frac{\omega\tau P}{d\sigma_0} - \frac{1}{\omega C_{ef}}\right] \quad (6)$$

From Eq. (6), the graphene array can be modeled as R-L-C circuit. $R$ and $L$ are due to the graphene elements. $C$ is induced by the space between graphene elements.

$$R = \frac{P}{d\sigma_0}, \quad L = \frac{\tau P}{d\sigma_0}, \text{ and } C = C_{ef} \quad (7)$$

The input impedance of the metal-backed substrate can be expressed as $Z_s = jZ_c \tan(\beta h)$ (8)
where $Z_c$ and $\beta$ are the characteristics impedance and wave number respectively.

The total impedance of the graphene MSF structure is

$$\frac{1}{Z_{in}} = \frac{1}{Z_g} + \frac{1}{Z_s} \quad (9)$$

The reflection coefficient $S_{11}$ of the graphene MSF structure is

$$S_{11} = \frac{Z_{in} - Z_0}{Z_{in} + Z_0} \quad (10)$$

where $Z_0$ is the free space impedance.

Absorption of graphene MSF structure is $A = 1 - |S_{11}|^2$

Absorption of the structure can also be expressed in terms of both reflection coefficient $S_{11}$ and transmission coefficient $S_{21}$ as $A = 1 - |S_{11}|^2 - |S_{21}|^2$

Perfect absorption is achieved when there is no reflection and no transmission.

### III. RESULTS AND DISCUSSION

The proposed graphene unit cell structure and the complete MSF structure are validated in CST Microwave Studio [16]. For the simulation of the unit cell, a Floquet port is applied to excite the impinging plane wave onto the unit cell.

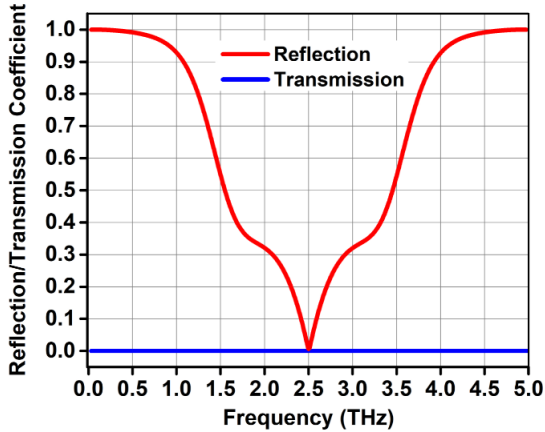

Fig. 2. Reflection and Transmission spectra of graphene MSF under normal incidence.

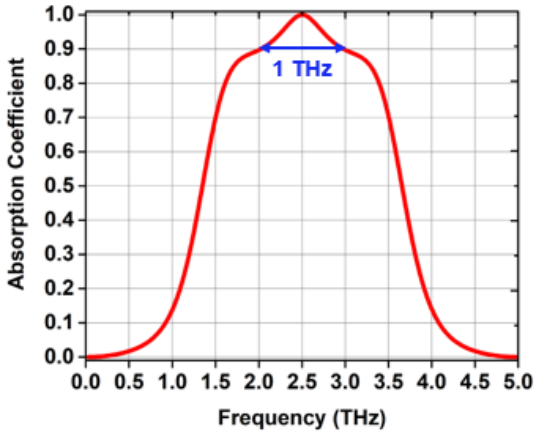

Fig. 3. Absorption spectra of graphene MSF under normal incidence.

The proposed graphene MSF provides 100% of absorption, zero reflection, and zero transmission at 2.5 THz. Reflection and transmission coefficient of the proposed MSF structure under normal incidence are shown in Fig. 2, whereas the absorption coefficient is shown in Fig. 3. From Fig. 3, we can observe that the graphene MSF provides a wide bandwidth of 1 THz (≈ 40%) ranging from 2.0 THz to 3.0 THz.

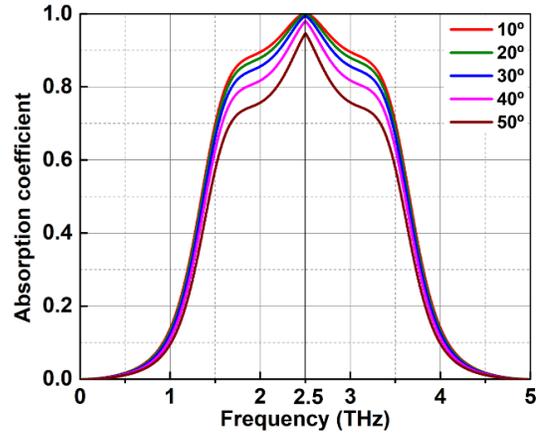

Fig. 4. Absorption spectra of graphene MSF under oblique incidence.

THz waves generally incident in different directions including normal incidence and oblique incidence. The maintenance of the perfect absorption performance and the central frequency at oblique incidence is the most important for a practical application. The absorption coefficients of the proposed graphene MSF structure at a different incidence angle from 0° to 50° are shown in Fig. 4. It can be noticed that the high absorption at central frequency for oblique incidences is more than 95%.

To investigate the reflection and absorption characteristics for both TE and TM polarization, we obtained the reflection spectra and the field distributions under normal incidence for both TE and TM polarization. Fig. 5 show the reflection spectra at normal incidence in both TE and TM polarizations. From this result, it is found that the proposed structure is insensitive to polarization. The reflection coefficient for both TE and TM polarization is −44dB (0.6%), which is very low. Ultimately, the proposed graphene THz MSF provides excellent absorption $(A = 1 - |S_{11}|^2)$, which is approximately 100% for both TE and TM polarization.

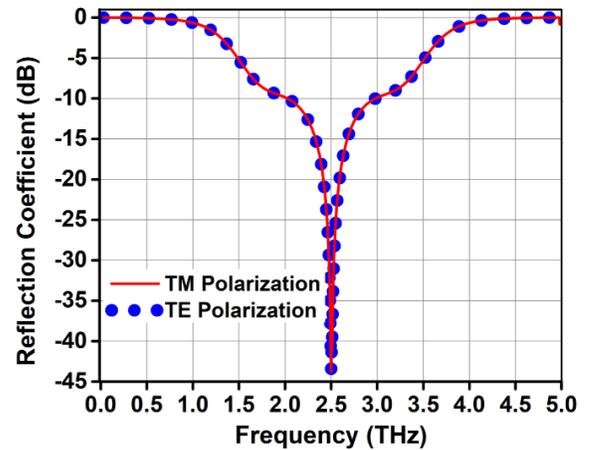

Fig. 5. (a) Reflection spectra for TE and TM polarization.

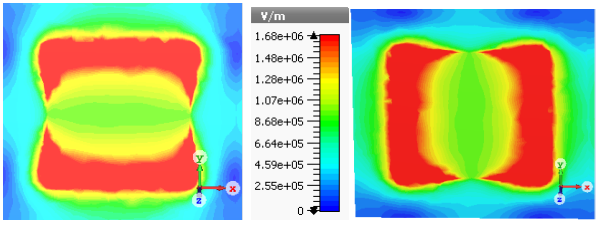

Fig. 6. Electric field distribution in (a) TE polarization and (b) TM polarization.

Fig. 6 displays the simulated electric field distributions of the proposed graphene MSF structure at 2.5 THz for both TE and TM polarizations. It can be remarked that the proposed structure exhibits strong electric field confinement around the graphene sheet on the top layer for the case of both TE and TM polarization, which leads to high THz absorption. The proposed graphene MSF structure provides wideband, excellent absorption, polarization-insensitive, larger insensitive oblique incidence. These characteristics have the potential for practical THz application.

In addition, the frequency reconfiguration of the graphene MSF is achieved by tuning the chemical potential of graphene. The chemical potential of graphene can be tuned in a wide range by means of electrical gating or chemical doping method. Fig. 7 shows the impact of tuning of the chemical potential of graphene on the reflection spectra and the frequency. Here, we consider the graphene chemical potential $\mu_c$ over the ranges from 0.5 eV to 0.6 eV. The small change of chemical potential makes a significant shift in the central frequency. The central frequency tuned from 2.5 THz to 3 THz as chemical potential $\mu_c$ increases from 0.5 eV to 0.6 eV. The minimum reflection at reconfigurable frequencies corresponds to maximum absorption. Fig. 7 also reveals that excellent absorption performance is maintained at reconfigurable frequencies.

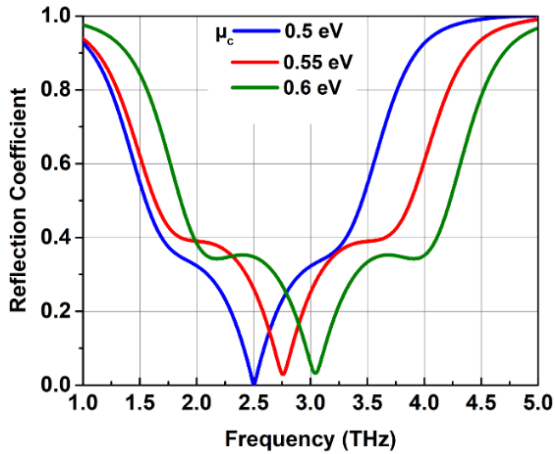

Fig. 7. Frequency reconfiguration of graphene THz MSF surface

## IV. CONCLUSION

In this work, we investigated a simple structured THz metasurface using graphene meta-atoms. An equivalent circuit model of the structure and its closed-form solution is presented. The graphene MSF thin structure provides 100% of absorption with wide bandwidth, zero reflection and zero transmission at 2.5 THz. In addition, the structure has the advantage of polarization-insensitive, frequency reconfiguration, and larger insensitive oblique incidence. The results reveal the effectiveness of the THz metasurface with graphene meta-atoms, which can be promising for THz wireless environment.

ACKNOWLEDGMENT

This research is supported by the European Union via the Horizon 2020: Future Emerging Topics - Research and Innovation Action call (FETOPEN-RIA), grant EU736876, project VISORSURF (http://www.visorsurf.eu).